\documentclass{iaus}
\usepackage{graphicx}

\title[Maser emission during post-AGB evolution] 
{Maser emission during post-AGB evolution}

\author[Desmurs, J.-F.]   
{Desmurs, J.-F.$^1$
}

\affiliation{$^1$Observatorio Astron\'omico Nacional, Madrid, Spain.\\ 
email: {\tt desmurs@oan.es}}

\pubyear{2012}
\volume{287}  
\pagerange{001}
\jname{Cosmic Masers - from OH to H{$_0$}}
\editors{Roy Booth, Liz Humphries \& Wouter Vlemmings Editor, eds.}

\begin{document}
\newcommand{\kms}{\mbox{km~s$^{-1}$}}

\maketitle

\begin{abstract}
This contribution reviews recent observational results concerning
astronomical masers toward post-AGB objects with a special attention to
water fountain sources and the prototypical source OH~231.8+4.2. These
sources represent a short transition phase in the evolution between
circumstellar envelopes around asymptotic giant branch stars and
planetary nebulae. The main masing species are considered and key
results are summarized.

\end{abstract}

\firstsection 
\section{Introduction}

After leaving the main sequence, stars of low and intermediate mass
travel across the Hertzsprung-Russell diagram and reach the asymptotic
giant branch (AGB). During this phase, the star ejects matter at a very
high rate (up to 10$^{-4}$ solar masses per year) in form of a slow (5
to 30~\kms), dense, isotropic wind. The resulting circumstellar
envelope often exhibits maser emission from several molecules, the most
common being SiO, H$_2$O and OH. These masers arise at different
distances from the star from different layers in the envelope, tracing
different physical and chemical conditions. SiO masers are found close
to the star (few stellar radii), then the water maser are a little
farther (a few hundreds of stellar radii) and farther out, the OH
masers (at a few thousands of stellar radii) see \cite {habing96}. As
the star follows its evolution to the planetary nebulae (PN) phase,
mass-loss stops and the envelope begins to ionized, such that the
masers emission disappear progressively. The SiO masers are supposed to
disappear first, the H$_2$O masers may survive a few hundreds of years
and OH masers remain for a period of $\sim$1000 years and can even be
found in PNe phase.

The evolution of the envelopes around AGB stars toward PNe, through the
phase of Proto Planetary Nebulae (pPNe) is yet poorly know, in
particular the shaping of PNe. While during the AGB phase the star
exhibits roughly spherical symmetry, about 10000 years later, PNe are
often asymmetrical (about 75\% see \cite [Manchado \etal\
2000]{manchado00}) showing axial symmetry, including multi-polar or
elliptical symmetry, and very collimated and fast jets. Bipolarity
appears very early in the post-AGB or pPN stage evolution.

\cite{sahai98} surveyed young PNe with the Hubble Space Telescope and
found that most of them were characterized by multi polar morphology
with collimated radial structures, and bright equatorial structures
indicating the presence of jets and disks/tori in some objects. They
propose that during the late AGB or early PPN stages the high-speed
collimated outflows carve out an imprint in the spherical AGB wind,
which provides the morphological signature for the development of
asymmetric PNe.

The mechanism explaining how axial symmetry appears during this
evolutionary phase is still an open question. Several models have been
proposed, that in general involve the interaction of very fast and
collimated flows, ejected by means of magneto-centrifugal launching,
with the AGB fossil shell. Interferometric observations of maser
emission in such sources allow us to get access to the formation and
evolution of these jets and with a very high spatial resolution.

This review will report meanly on recent results published since the
previous IAU 242 maser symposium, held in Alice Springs in 2007.

\section {Surveys}

\begin{figure}[h]
\begin{center}
 \includegraphics[width=3.4in]{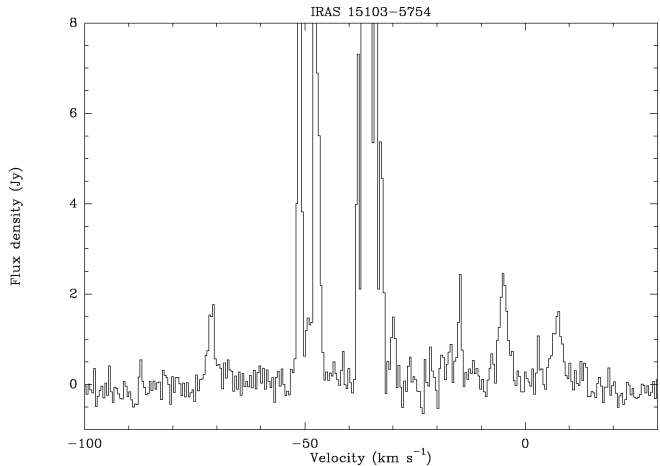} 
 \caption{IRAS 15103-5754, first water fountain detected in a planetary
 nebula (\cite[Su\'arez \etal\ 2009]{suarez09}).}
   \label{fig1}
\end{center}
\end{figure}

During the last years, several surveys have been conducted to discover
new maser emission toward post-AGB objects. \cite{deacon07}, searched
for water masers (at 22~GHz) and SiO masers (at 86~GHz) using
respectively the telescopes of Tibdinbilla-70m and Mopra-22m. They
observed a list of 85 sources in total (11 of them in SiO), selected on
the basis of their OH 1612~MHz spectra, and get 21 detections (3 in
SiO), out of which 5 sources present high velocity profiles and one
source show a wide double peak profile of the SiO maser.
\cite[Su\'arez \etal\ (2007, 2009)]{suarez07, suarez09}, searched for
water maser in the northern hemisphere (with the Robledo-70m antenna)
and in the southern hemisphere (Parkes-64m). They surveyed 179 sources
(mostly pPNe \& PNe) and detected 9 sources (4 pPNe and 5 PNe, one of
these, IRAS~15103-5754, has water fountain characteristics, see Figure
\ref{fig1})

Using the 100 m telescope of Effelsberg, \cite {desmurs10} undertook a
high sensitivity discrete source survey for the first excited state of
OH maser emission (J = 5/2, $^2\Pi_{3/2}$ at 6 GHz) in the direction of
65 PNe and pPNe exhibiting 18 cm OH emission (main and/or satellite
lines). They detected two sources at 6035 MHz (5 cm), both of them
young (or very young) PNe.  And very recently, \cite{amiri12},
conducted a sensitive survey with the Effelsberg antenna of water maser
emission toward 74 post-AGBs, and found 6 new water fountain candidates
showing a double peak profile.

\section {Water Fountains}

\begin{figure}[h]
\begin{center}
 \includegraphics[width=\textwidth]{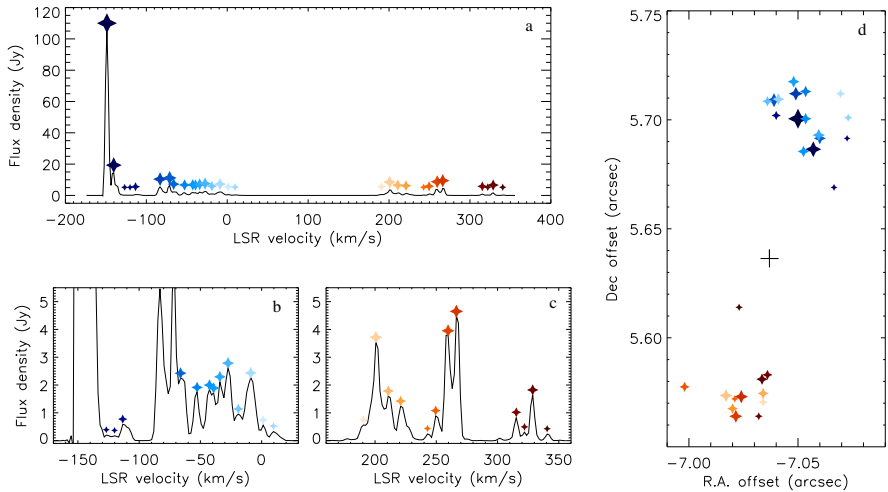} 
 \caption{e-VLA spectrum and map of the water fountain IRAS 18113-2503,
 see \cite {gomez11} for details.}
   \label{fig2}
\end{center}
\end{figure}

In the class of the proto-planetary nebulae, there is a very
interesting sub-class of young pPNe called the water fountains. They
present both hydroxyl and water maser emission, however their
characteristics differ from those typical of AGB stars\footnote{OH
masers in AGB star generally exhibit double-peaked profiles covering up
to 25 km/s (e.g. \cite[te Lintel \etal\ 1989]{telintel89}), and H$_2$O
maser spectra present a velocity range within the OH maser one and
their profiles are more irregular.}.  First of all, H$_2$O and OH maser
spectra exhibit a double peaked profile with the peaks symmetrically
distributed about the star velocity.  But, unlike in AGB stars, H$_2$O
maser are spread over a larger velocity range than the OH masers (see
\cite[Imai \etal\ 2008]{imai08} for example) and higher velocity (up to
400~\kms~ see Figure \ref{fig2}) than OH maser velocity or AGB radial
expansion wind (10-20~\kms~ \cite[te Lintel \etal\ 1989]{telintel89}).
High spatial resolution observations of the water masers emission in
these objects reveals bipolar distribution and highly collimated
outflows (hence the name of water fountain).  The first member of this
subclass of pPN, W~43A, was first observed by \cite{imai02} with the
VLBA, and showed a well collimated and precessing jet with an outflow
velocity of $\sim$145~\kms. When optical images are available, the
masers appear coincident with the optical bipolar structures (see for
example IRAS~16342-3814, \cite [Claussen \etal\ 2009]{claussen09} and
this proceeding).

\begin{table}\def~{\hphantom{0}}
  \begin{center}
  \caption{Confirmed water fountains$^a$ }
  \label{tab1}
   \begin{tabular}{lcccc}\hline
     PNe          & Other name  & OH Velocity  & H$_2$O Velocity&References\\
                  &             & range & range&\\
                  &             &(in \kms)&(in \kms)&\\
\hline
 IRAS 15445-5449 &OH~326.5-0.4&&90&\cite{deacon07}\\
 IRAS 15544-5332 &OH~325.8-0.3&&74& \cite{deacon07}\\
 IRAS 16342-3814 &OH~344.1+5.8&&260& \cite{claussen09}\\
 IRAS 16552-3050 &GLMP 498&&&\cite {suarez07}\\
 IRAS 18043-2116 &OH~0.9-0.4&&400&\cite {walsh09}\\
 IRAS 18113-2503 &PM~1-221 &&500&\cite {gomez11}\\
 IRAS 18139-1816 &OH~12.8-0.9&23&42&\cite {boboltz07}\\
 IRAS 18286-0959 &OH~21.79-0.1&&200&\cite {yung11}\\
 IRAS 18450-0148 &W~43A/OH~31.0+0.0 &&180&\cite {imai02} \\
 IRAS 18460-0151 &OH~31.0-0.2&20&300& \cite {imai08}\\
 IRAS 18596+0315 &OH~37.1-0.8&30&60&\cite {amiri11}\\
 IRAS 19067+0811 &OH~42.3-0.1&20&70&\cite{gomez94}\\
 IRAS 19134+2131 &G054.8+4.6 &&105&\cite {imai07}\\ 
 IRAS 19190+1102 &PM~1-298&&100&\cite {day10}\\
 IRAS 15103-5754 &G320.9-0.2 &&80&\cite {suarez09}\\
\hline
\end{tabular}
\begin{tabular}{l}
\noindent\scriptsize{$^a$ +6 new water fountains candidates (see
  \cite[Amiri \etal\ 2012]{amiri12}).}
\end{tabular}
 \end{center}
\end{table}

Table \ref{tab1} shows a list of the confirmed\footnote{The candidates
sources IRAS 07331+0021 and IRAS 13500-6106 from \cite{suarez09} turned
to be ``classical'' proto-planetary nebulae and not water fountain.}
water fountains. Up to now, 14 sources have been identified as water
fountains and 6 new sources have been recently found by \cite{amiri12}
and are good candidates to be classified as such (they all present a
double peak spectra and high velocity profiles).  The most recent
source confirmed to belong to this sub-class is IRAS 18113-2503
(\cite[G\'omez \etal\ 2011]{gomez11}, see Figure \ref{fig2}). The
source shows the typical double peak spectra, with a very large
velocity dispersion and the peaks of water emission are separated by
about 500~\kms~ (from -150 to +350~\kms~ LSR). It is likely to be the
fastest outflows observed up to now with a velocity of at least
250~\kms. The e-VLA map clearly shows a bipolar spatial distribution
with a blueshifted part to the north and a redshited lobe to the south.
\\ IRAS 19067+0811 is also an interesting source, observations in 1988
clearly detected a double peak H$_2$O spectrum covering a velocity
range of twice of the velocity range of the OH spectra, but \cite
{gomez94} only detected OH maser emission. And finally, I would like to
mention IRAS 15103-5754 a very peculiar source as it is the first
confirmed water fountain that is not a pPN but a PN (see \cite[Su\'arez
\etal\ 2009]{suarez09} and these proceedings).

\subsection {Outflows Proper motion}

A careful analysis of the 3D structure of jets (including proper
motions studies) is a fundamental element for the development of
theoretical models of the PN shaping mechanisms and of the nature of
the outflows that give rise to the complex structure found in many pPNe
and PNe.

\begin{figure}[h]
\begin{center}
 \includegraphics[width=3.4in]{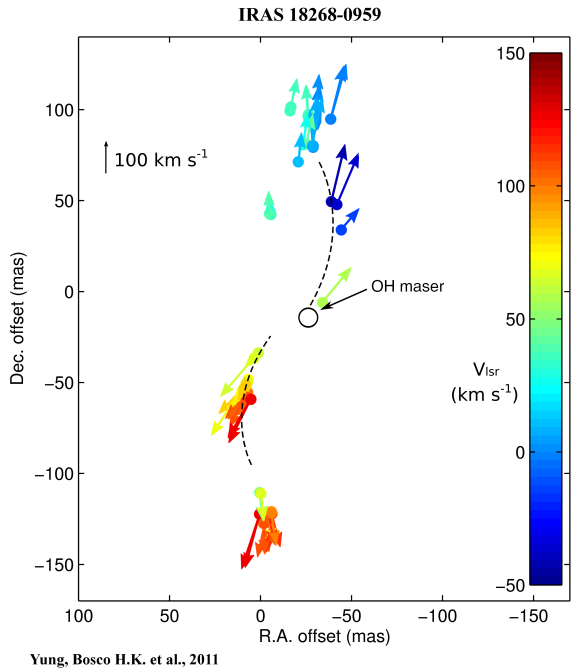} 
 \caption{Proper motion of H$_2$O maser features measured in IRAS
 18286-0959 (from \cite [Yung \etal\ 2011]{yung11})}
   \label{fig3}
\end{center}
\end{figure}

In W~43A, VLBA observations of H$_2$O maser emission show very well
collimated jets with a velocity of the order of $\sim$145~\kms~
(\cite[Imai \etal\ 2002, 2005]{imai02, imai05}). The proper motion
analysis also demonstrates that the jets exhibit a spiral pattern and
is precessing with a period of 55 years.
In the source IRAS~16342-3814 (e.g. \cite[Claussen \etal\
2009]{claussen09} and these proceedings) found that water masers lie on
opposite sides of the optical nebula and their distribution is
generally tangential to the inferred jet axis. Proper-motion
measurements give a velocity of the jet at the position of the extreme
velocity maser components of at least 155 to 180~\kms~ but no direct
evidence for precession was found, maser features appear to follow a
purely radial motion (no curve trajectories are observed).
A detailed morpho-kinematical structure analysis of the H$_2$O masers
from the water fountain IRAS 18286-0959 has been carried on by \cite
{yung11} (see Figure \ref{fig3}). Observations are best interpreted by
a model with two precessing jets (or ``double helix'' outflow pattern)
with velocities of up to 138~\kms~ and a precessing period of less than
60 years.

This proper motion studies allow also to estimate other parameters like
the dynamical age of these outflows and they are found to be
surprisingly young, of the order of few tens of years, up to 150 years,
which suggests that the evolutionary stage that these water-fountain
sources represent is likely to be very short: 50 years for W~43A (see
\cite[Imai \etal\ 2005]{imai05}), $\sim$30 years for IRAS 18286-0959
(\cite [Yung \etal\ 2011]{yung11}), $\sim$59 years for IRAS 19190+1102
(\cite [Day \etal\ 2010]{day10}) and about 120 years for IRAS
16342-3814 (\cite [Claussen \etal\ 2009]{claussen09}).  VLBI astrometry
of H$_2$O masers also allow to derive more accurate distance of these
sources.

\subsection {Polarization and magnetic collimation}

The origin of the jet collimation is still an open question but several
models (\cite[Chevalier \& Luo 1994]{chevalier94},
\cite[Garc\'{\i}a-Segura \etal\ 1999]{garcia99}) have shown that the
magnetic field could be a dominant factor in jet collimation
(\cite[Blackman \etal\ 2001]{blackman01}, \cite[Garc\'{\i}a-Segura
\etal\ 2005]{garcia05}) in post-AGB stars.  The Zeeman effect produces
a shift in frequency between the two circular polarizations (LCP and
RCP) that is directly proportional to the strength of the magnetic
field (projected on the line of sight). Hence, by measuring this shift,
we can deduce the value of the magnetic field. Several molecules giving
rise to maser emission are sensible to this effect (like OH and
H$_2$O), and then provide a unique tool for studying the role of the
magnetic field in the jet collimation (of water fountains for
example).\\ Full polarization observations measuring both linear and
circular polarization toward the archetype water fountain W~43A have
been conducted (see \cite[Vlemmings \etal\ 2006]{vlemmings06},
\cite[Amiri \etal\ 2010]{amiri10}). A strong toroidal magnetic field
has been measured, with an estimated strength on the surface of the
star as high as 1.6~G (see \cite[Vlemmings \etal\
2006]{vlemmings06}). Such a magnetic field is strong enough to actively
participate in the collimation of the jets.

Recently, \cite{wolak11} published a single dish survey conducted with
the Nancay radio telescope toward 152 late type stars, out of which 24
were post AGB sources. In more than 75\% of the sample, they detected
polarization features and a magnetic field strength of 0.3 to
2.3~mG. In summary, strong magnetic field are observed, strong enough
to play a major role in shaping and driving the outflows in water
fountains.

\section{SiO PPN}
SiO maser are very rare in pPNe, very few sources have been detected,
OH~15.7+0.8 (tentative detection), OH~19.2-1.0, W~43A, OH~42.3-0.1,
IRAS~15452-5459, IRAS~19312+1950 and OH~231.8+4.2. The last one
discovered is IRAS~15452-5459 (\cite[Deacon \etal\ 2007]{deacon07}).
The SiO maser emission of two of these sources has been mapped, W~43A
and OH~231.8+4.2.  In the first case, the spatial distribution was
found to be compatible with a bi-conical decelerating flow and in the
case of OH~231.8+4.2, the distribution can be described by a torus with
rotation and infalling velocities.
 of the same order and within a range
between $\sim$\,7 and $\sim$10~\kms~ (see \cite[S\'anchez Contreras
\etal\ 2002]{sanchez02}).

\section{The prototype of bipolar proto Planetary Nebulae: OH~231.8+4.2}

\begin{figure}[h]
\begin{center}
 \includegraphics[width=\textwidth]{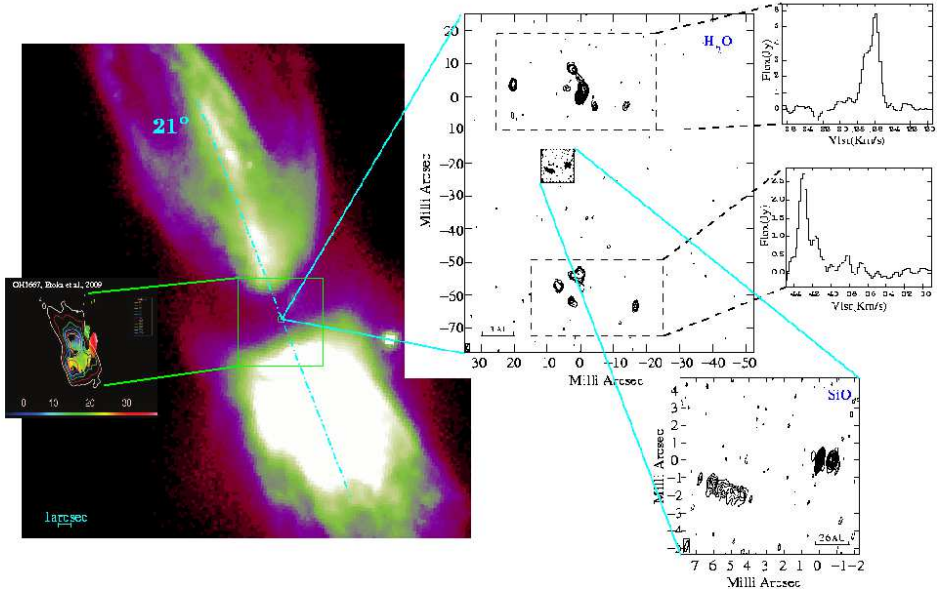} 
 \caption{Composition image summarizing interferometric observations OH,
 H$_2$O and SiO toward OH~231.8+4.2}
   \label{fig4}
\end{center}
\end{figure}

OH 231.8+4.2 (also known as the Calabash nebula or QX Pup) is well
studied and prototype of bipolar pPNe.  It is located in the open
cluster M46 at a distance of 1.5 kpc $\pm$0.05 (\cite [Choi \etal\
2012]{choi12}) and the inclination of the bipolar axis with respect to
the plane of the sky, $\sim 36^o$.. The central source is a binary
system formed by an M9-10 III Mira variable (i.e. an AGB star) and an
A0 main sequence companion, as revealed from optical spectroscopy by
\cite{sanchez04}. This remarkable bipolar nebula shows all the signs of
post-AGB evolution: fast bipolar outflows with velocities
$\sim$200--400~\kms, shock-excited gas and shock-induced chemistry.
Mid-infrared MIDI observations (\cite[Matsuura \etal\
2006]{matsuura06}) show the presence of a compact circumstellar region
with an inner radius of 40-50 AU. An equatorial torus is observed at
distances greater than 1 arcsec, however, no trace of rotation is found
at this scale and the gas is in expansion, as shown by CO and OH
emission data (\cite[Alcolea \etal\ 2001]{alcolea01}, \cite[Zijlstra
\etal\ 2001]{zijlstra01}).  Hubble space telescope observations clearly
show two extended lobes and PdBI CO observations measured a traveling
speed of the molecular outflow of the order of 400~\kms.

OH~231.8+4.2 still shows intense SiO masers, contrarily to what happens
in the majority of pPNe.  The SiO maser emission arises from several
compact, bright spots forming a structure elongated in the direction
perpendicular to the symmetry axis of the nebula.

Figure \ref{fig4} is a composition image summarizing OH, H$_2$O and SiO
maser emission observations compared with the HST image of the nebula
(taken with the WFPC2 \cite[Bujarrabal \etal\ 2002]{bujarrabal02}). The
left panel presents the velocity map of the OH maser emission at
1667-MHz aligned over the L-band image obtained at the VLT by \cite
{matsuura06}. Top right panels show the total intensity map of the
H$_2$O maser for the two main regions.  The small square map at bottom
right indicates the position of the map of SiO maser obtained by
\cite{sanchez02}.
OH~231.8+4.2 is a strong emitter in the OH ground state line at 1667
MHz. This strong maser emission, radiated by the circumstellar material
around OH~231.8+4.2, was mapped with MERLIN by \cite{etoka09} The OH
maser distribution (4~arcsecond) traces a ring-like structure
presenting a velocity gradient that is explained by the authors as the
blueshifted rim of the bi-conical outflow.  The distribution of the
polarization vectors associated with the maser spots attests a
well-organized magnetic field which seems to be flaring out in the same
direction as the outflow.
H$_2$O maser emission is distributed in two distinct regions of
$\sim$20 mas in size, spatially displaced by 60 milli-arcs (less than
100~AU, comparable to the size of the AGB envelopes) along an axis
oriented nearly north-south, similarly to the axis of the optical
nebula.  The expansion velocity of the H$_2$O masers spots is very low
compared to water fountain jets and lower than that of the OH maser
spots. Proper motion observations (\cite[Leal-Ferreira \etal\
2012]{leal12}, \cite [Desmurs \etal\ 2012]{desmurs12}) derived
velocities on the sky of the order of 2--3 mas/year. Taking into
account the inclination angle of the source, this corresponds to an
average separation velocity of 15~\kms. Moreover, the H$_2$O emission
is not as well collimated as in water-fountains. Linear polarization of
H$_2$O maser yields a value of the magnetic field, assuming a toroidal
structure, of 1.5--2.0~G on the stellar surface (see
\cite[Leal-Ferreira \etal\ 2012]{leal12}).
SiO masers are tentatively found to be placed between the two H$_2$O
maser emitting regions and rise from several compact features tracing
an elongated structure in the direction perpendicular to the symmetry
axis of the nebula. Probably a disk rotating around the M-type
star. The distribution is consistent with an equatorial torus with a
radius of $\sim$\,6~AU around the central star. A complex velocity
gradient was found along the torus, which suggests rotation and infall
of material towards the star with velocities of the same order and
within a range between $\sim$\,7 and $\sim$10~\kms~ (see
\cite[S\'anchez Contreras \etal\ 2002]{sanchez02}).

\begin{acknowledgments}
I would like to acknowledge financial support from the Visiting
Scientist grant from the National Research Foundation of South Africa.
I would like also to acknowledge V.Bujarrabal for his very useful
comments and careful reading of the manuscript. Thank's POD!
\end{acknowledgments}



\begin{thebibliography}{}

\bibitem[Alcolea \etal\ (2001)]{alcolea01}
        {Alcolea, J., Bujarrabal, V., S\'anchez Contreras, C., Neri, R.
        \& Zweigle, J.} 2001, \textit{A\&A} 373, 932.
\bibitem[Amiri \etal\ (2010)]{amiri10}
        {Amiri N., Vlemmings W., van Langevelde H.J.} 2010, 
        \textit{A\&A}, 509, 26.
\bibitem[Amiri \etal\ (2011)]{amiri11}
        {Amiri N., Vlemmings W., van Langevelde H.J.} 2011, 
        \textit{A\&A}, 532, 149. 
\bibitem[Amiri \etal\ (2012)]{amiri12}
        {Amiri, N, Vlemmings, W.H.T. \& van Langevelde, H.J.} 2012,
        \textit{A\&A in preparation}
\bibitem[Boboltz \& Marvel (2007)]{boboltz07}
        {Boboltz, D.A. \& Marvel, K.B.} 2007, \textit{ApJ}, 665, 680.
\bibitem[Blackman \etal\ (2001)]{blackman01}
        {Blackman, E.G., Frank, A., Markiel, J.A., Thomas, J.H. \& Van
        Horn, H. M.} 2001,  \textit{Nature} 409, 485.
\bibitem[Bujarrabal \etal\ (2002)]{bujarrabal02}	
	{Bujarrabal, V., Alcolea, J., S\'anchez Contreras, C. \& Sahai,
	R.} 2002, \textit{A\&A}, 389, 271.
\bibitem[Chevalier \& Luo (1994)]{chevalier94}
        {Chevalier, R.A. \& Luo, D.} 1994,\textit{ApJ} 421, 225.
\bibitem[Choi \etal\ (2012)]{choi12}
        {Choi, Y.K. et al.} 2012, \textit{this proceeding}
\bibitem[Claussen \etal\ (2009)]{claussen09}	
	{Claussen, M.J., Sahai, R. \& Morris, M.R.} 2009, \textit{ApJ},
	691, 219.
\bibitem[Day \etal\ (2010)]{day10}
        {Day, F.M., Pihlström, Y.M., Claussen, M.J. \& Sahai, R.} 2010,
        \textit{ApJ}, 713, 986.
\bibitem[Deacon \etal\ (2007)]{deacon07}
        {Deacon, R.M., Chapman, J.M., Green, A.J. \& Sevenster, M.N.}
        2007, \textit{ApJ}, 658, 1096.
\bibitem[Desmurs \etal\ (2010)]{desmurs10}
        {Desmurs, J.-F., Baudry, A., Sivagnanam, P., Henkel, C.,
        Richards, A. M. S. \& Bains, I.} 2010, \textit{A\&A}, 520, 45.
\bibitem[Desmurs \etal\ (2012)]{desmurs12}
        {Desmurs, J.-F. \etal\ } 2012  \textit{A\&A in preparation}
\bibitem[Etoka \etal\ (2009)]{etoka09}
        {Etoka, S., Zijlstra, A., Richards, A.M., Matsuura, M. \& 
        Lagadec, E.} 2009 \textit{ASPC} 404, 311.
\bibitem[Garc\'{\i}a-Segura \etal\ (1999)]{garcia99}
        {Garc\'{\i}a-Segura, G., Langer, N., R\'ozuczka, M. \& Franco,
        J.} 1999, \textit{ApJ}, 517, 767. 
\bibitem[Garc\'{\i}a-Segura \etal\ (2005)]{garcia05}
        {Garc\'{\i}a-Segura, G., L\'opez, J. A. \& Franco, J.} 2005,
        \textit{ApJ}, 618, 919.
\bibitem[G\'omez \etal\ (1994)]{gomez94}
        {G\'omez, Y., Rodr\'iguez, L.F., Contreras, M.E. \& Moran, J. 
        M.}, 1994, \textit{RMxAA},28, 97.
\bibitem[G\'omez \etal\ (2011)]{gomez11}
        {G\'omez, J.F., Rizzo, R.J., Suarez, O. \& Ñiranda, L.F.} 2011,
        \textit{ApJL}, 739, L14.
\bibitem[Habing (1996)]{habing96}
        {Habing, H. J.} 1996, \textit{A\&A Rev.}, 7, 97.
\bibitem[Imai \etal\ (2002)]{imai02}
        {Imai, H., Obara, K., Diamond, P.J., Omodaka, T. \& Sasao, T.}
        2002, \textit{Nature} 417, 829.
\bibitem[Imai \etal\ (2005)]{imai05}
        {Imai, H., Nakashima, J.I., Diamond, P.J., Miyazaki, A. \&
        Deguchi, S.} 2005, \textit{ApJ} 622, L125.
\bibitem[Imai \etal\ (2007)]{imai07}
        {Imai, H., Sahai, R. \& Morris, M} 2007, \textit{ApJ}, 669, 424.
\bibitem[Imai \etal\ (2008)]{imai08} {Imai, H., Diamond, P., Nakashima,
        J.I., Kwok, S., \& Deguchi, S.} 2008, \textit{Proceedings of
        the 9th European VLBI Network Symposium on The role of VLBI in
        the Golden Age for Radio Astronomy and EVN Users
        Meeting. September 23-26, 2008. Bologna, Italy}, 60
\bibitem[Imai \etal\ (2009)]{imai09}
        {Imai} 2009, \textit{}
\bibitem[Leal-Ferreira \etal\ (2012)]{leal12}
        {Leal-Ferreira, M.L., Vlemmings, W.H.T., Diamond, P.J.,
        Kemball, A., Amiri, N. \& Desmurs, J.-F.} 2012, \textit{A\&A
        accepted (arXiv:1201.3839v1)} 
\bibitem[Manchado \etal\ (2000)]{manchado00} 
        {Manchado, A., Villaver, E., Stanghellini, L., \& Guerrero,
        M. A.} 2000, \textit{in ASP Conf. Ser. 199, Asymmetrical
        Planetary Nebulae II: From Origins to Microstructures,
        ed. J. H. Kastner et al. (San Francisco: ASP}, 17.
\bibitem[Matsuura \etal\ (2006)]{matsuura06}
        {Matsuura, M., Chesneau, O., Zijlstra, A.A., Jaffe, W., Waters,
        L.B.F.M., Yates, J.A., Lagadec, E.; Gledhill, T., Etoka, S. \&
        Richards, A.M.S.} 2006, \textit{ApJ} 646, 123.
\bibitem[Sahai \& Trauger (1998)]{sahai98}
        {Sahai, R., \& Trauger, J. T.} 1998, \textit{AJ}, 116, 1357
\bibitem[S\'anchez Contreras \etal\ (2002)]{sanchez02}
        {S\'anchez Contreras, C., Desmurs, J.-F., Bujarrabal, V.,
        Alcolea, J., \& Colomer,F.} 2002, \textit{A\&A}, 385, L1.
\bibitem[S\'anchez Contreras \etal\ (2004)]{sanchez04}
        {S\'anchez Contreras, C., Gil de Paz, A., Sahai, R.} 2004,
        \textit{ApJ}, 616, 519.
\bibitem[Su\'arez \etal\ (2007)]{suarez07}
        {Su\'arez, O., G\'omez, J.F. \& Morata, O.} 2007, \textit{A\&A}
        467, 1085.
\bibitem[Su\'arez \etal\ (2008)]{suarez08}
        {Su\'arez, O., G\'omez, J.F. \& Miranda, L.F.} 2007,
        \textit{ApJ} 689, 430.
\bibitem[Su\'arez \etal\ (2009)]{suarez09}
        {Su\'arez, O., G\'omez, J.F., Miranda, L.F., Torrelles, J.M.;
        G\'omez, Y., Anglada, G. \& Morata, O.} 2009, \textit{A\&A}
        505, 217.
\bibitem[te Lintel \etal\ (1989)]{telintel89}
        {te Lintel Hekkert, P., Versteege-Hensel, H.A., Habing, H.J. \&
        Wiertz, M.} 1989, \textit{A\&AS}, 78, 399.
\bibitem[Vlemmings \etal\ (2006)]{vlemmings06}
        {Vlemmings, W.H.T., Diamond, P.J. \& Imai, H.}\textit{Nature}
        440, 58.
\bibitem[Walsh \etal\ (2009)]{walsh09}
	{Walsh, A.J., Breen, S.L., Bains, I. \& Vlemmings, W.H.T.},
	2009, \textit{MNRAS}, 394, 70.
\bibitem[Wolak \etal\ (2011)]{wolak11}
        {Wolak, P., Szymczak, M. \& Gerard, E.} 2011, \textit{A\&A} 
\bibitem[Yung \etal\ (2011)]{yung11}
        {Yung, Bosco H. K., Nakashima, J., Imai, H., Deguchi, S.,
        Diamond, P.J. \& Kwok, S.} 2011, \textit{ApJ}, 741, 94
\bibitem[Zijlstra \etal\ (2001)]{zijlstra01}
        {Zijlstra, A., Chapman, J.M., te Lintel Hekkert, P., Likkel,
        L., Comeron, F., Norris, R.P., Molster, F.J. \& Cohen, R. J.}
        2001, \textit{MNRAS} 322, 280.
\end{thebibliography}
\end{document}